\documentclass[11pt]{article}

\usepackage[utf8]{inputenc}
\usepackage[T1]{fontenc}
\usepackage[margin=1in]{geometry}
\usepackage{graphicx}
\usepackage{amsmath,amssymb}
\usepackage{siunitx}
\usepackage{booktabs}
\usepackage{xcolor}
\usepackage{authblk}
\usepackage[font=small,labelfont=bf]{caption}
\usepackage{placeins}
\usepackage{float}
\usepackage[numbers,sort&compress]{natbib}
\usepackage[colorlinks=true,allcolors=blue!55!black]{hyperref}

\DeclareSIUnit{\micrometre}{\ensuremath{\mu\mathrm{m}}}
\newcommand{\aL}{\ensuremath{\alpha\!\cdot\!L}}
\newcommand{\Ca}{\ensuremath{^{40}\mathrm{Ca}^{+}}}

\title{\bfseries One Aperture, Two Colors:\\
Depth-Allocated Bilayer Silicon Nitride Gratings for Trapped-Ion Addressing}

\author[1]{Gyanendra~Yadav\thanks{Corresponding author:
\texttt{gyadav@nus.edu.sg}.}}
\affil[1]{Centre for Quantum Technologies, National University of Singapore, Singapore}
\date{}

\begin{document}
\maketitle


\vspace{0.3em}
\begin{abstract}
Scaling multicolor quantum control on chip presents a geometric conflict: each added wavelength 
traditionally requires another routing corridor, emitter, and electrode opening, consuming 
trap-surface area and exposing charge-susceptible dielectric near the quantum target. This work 
introduces an all-planar, vertically stacked silicon-nitride architecture that co-registers 
independently fed optical transitions within one $24\times\SI{24}{\micrometre}$ footprint. 
For \Ca{}, the device focuses the \SI{729.4}{\nano\meter} qubit and \SI{854.2}{\nano\meter} repump 
wavelengths onto a shared site \SI{70}{\micrometre} above the upper film. A hierarchical design 
combines a differentiable cell search for a three-level scatterer, analytic confocal phase synthesis, 
and deterministic depth allocation across two etches per layer, preserving an explicit 
\SI{125}{\nano\meter} propagation-axis line-and-space floor without grayscale patterning. 
Retargeting the upper-layer phase co-registers the focal maxima to within \SI{0.1}{\micrometre} in 
mesh-converged Tidy3D and independently constructed Lumerical FDTD models, retaining more than 99.9\% 
of each color's peak intensity at the common coordinate with focusing efficiencies of 0.410 and 0.168. 
At \SI{5}{\micrometre} ion spacing, the qubit channel reaches $-24.8$~dB point crosstalk, remains 
below $-18.0$~dB across a \SI{0.5}{\micrometre}-radius positioning disk, and yields $-23.0$~dB 
symmetric-traceless field-gradient suppression. In the analytic House conductor-plane geometry, 
the shared footprint provides \SI{15.3}{\micrometre} clearance to the nearest radio-frequency rail 
with zero RF-null displacement. By uniting separate optical inputs and transition-specific 
wavefronts within a single ion-facing aperture, this design hierarchy establishes a transferable 
template for addressing other atomic wavelength pairs and dense multicolor quantum interfaces.
\end{abstract}

\section{Introduction}

Scaling a quantum processor requires more than increasing qubit count: optical control must reach
many sites without consuming the surfaces, routing corridors, and alignment degrees of freedom
needed by the processor itself. This constraint is acute in surface-electrode ion traps and
multizone quantum charge-coupled-device (QCCD) architectures, where dense radio-frequency (RF) and
control electrodes share a two-dimensional plane with every optical opening
~\cite{Mordini2025}. When each
wavelength is assigned a separate emitter, adding optical functions increases
the ion-facing footprint and can enlarge the exposed dielectric area.
Illumination of these dielectrics can induce charging, stray electric fields,
and excess micromotion
~\cite{Harlander2010,Niffenegger2020,Ivory2021}. These constraints motivate an optical interface that delivers independently
controlled wavelengths through a shared aperture, allowing spectral
functionality to grow without a corresponding increase in electrode openings.

The \Ca{} system makes this architectural problem concrete. Its narrow
$4^2S_{1/2}\!\leftrightarrow\!3^2D_{5/2}$ electric-quadrupole transition at
\SI{729.4}{\nano\meter} supports coherent qubit rotations and resolved-sideband cooling, whereas
\SI{854.2}{\nano\meter} couples $3^2D_{5/2}$ to $4^2P_{3/2}$ for quenching, repumping, and state
reset~\cite{Schindler2013,RicciVasquez2023}. These distinct functions motivate
separate optical inputs and wavelength-specific wavefronts delivered to a
common site. Free-space optics can combine the two colors, but their relative
pointing and focus must be aligned and maintained. Integrated delivery offers
a route to encoding their nominal registration in the device geometry.
Both wavelengths lie in a low-loss window of silicon nitride and are accessible
with mature laser sources~\cite{Poon2024}.

Trap-integrated photonics has demonstrated single-wavelength focusing, multicolor delivery from
laterally distributed emitters, shared broadband gratings, and multizone QCCD control
~\cite{Mehta2016,Mehta2020,Niffenegger2020,Shirao2022,Klawson2022,Mordini2025}.
Multimode photonic circuits can address several ions through one grating, while a wafer-scale hybrid
waveguide-lens/three-dimensional-printed micromirror interface has demonstrated broadband
405--\SI{880}{\nano\meter} delivery~\cite{Momenzadeh2026,Klawson2026}. These broadband common
optics are complementary to the problem considered here: they maximize spectral reach through a
shared optical train, whereas color-specific gratings can synthesize different narrowband
wavefronts. Multi-band gratings and waveguide holograms have likewise directed several wavelengths
toward a common free-space region~\cite{Zhou2018,DeVocht2026}, and bilayer or multilayer gratings
have enhanced coupling or controlled independent waves
~\cite{Mak2018,Vitali2022,Ashtiani2021,DiCroce2026,LiMiller2026}. These results establish shared
broadband transport, multiwavelength beam delivery, and multilayer emission separately. The
distinct capability investigated here is an all-planar, independently fed pair of vertically stacked
narrowband gratings whose separately synthesized wavefronts co-register at one microscopic target.
Once coupled into the on-chip routes, such an interface fixes their relative focus in the masks and
removes independent downstream beam steering from the two-color alignment problem.

That capability is realized with a hierarchical design method that uses optimization only where it
adds physical leverage. A differentiable single-film calculation searches the nonlinear local
scattering response of a three-level cell; the known aperture-scale confocal phase is synthesized
analytically; and deterministic depth allocation encodes the radiation envelope in two discrete
etch depths per layer. The method connects fabrication-aware inverse design in couplers,
metaoptics, and free-space emitters
~\cite{Piggott2017,Schubert2022,Carfagno2023,RoquesCarmes2022,Yulaev2023,Hermann2022} to an
interpretable multilayer aperture. Unlike a weak-scattering duty taper that can drive teeth or gaps
below the process floor, the final four etch masks preserve a stated
\SI{125}{\nano\meter} propagation-axis line-and-space floor without grayscale lithography
~\cite{Benedikovic2014,Benedikovic2017,Chen2017,Li2013,Luo2022,LiMiller2026}.

\begin{figure}[htbp]
  \centering
  \includegraphics[width=\linewidth]{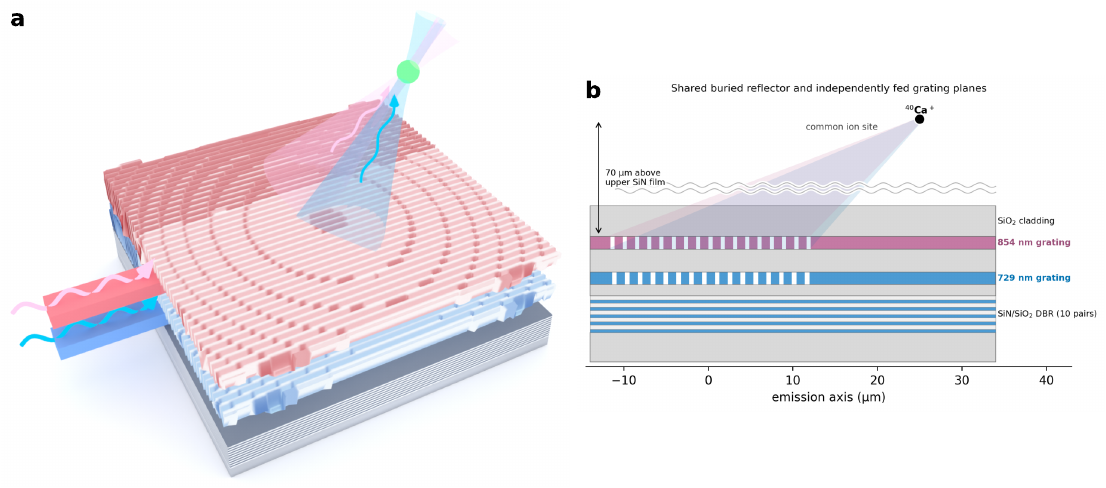}
  \caption{\textbf{Shared-aperture bilayer focusing architecture.}
  (a) Physically based rendering generated from the final phase-registered height maps. The
  two independently fed SiN layers form confocal, depth-allocated apertures whose beams
  converge on one \Ca{} site. Vertical relief is exaggerated for visibility; the simulated
  aperture is square. (b) Physical stack used in the full-stack solvers: a ten-pair SiN/SiO$_2$
  DBR, the lower \SI{729}{\nano\meter} grating, a \SI{1}{\micrometre} oxide interlayer, and
  the upper \SI{854}{\nano\meter} grating under oxide cladding. Both colors propagate to a
  common ion plane \SI{70}{\micrometre} above the upper SiN film.}
  \label{fig:arch}
\end{figure}

The resulting device uses two independently fed \SI{200}{\nano\meter}
silicon-nitride (SiN) gratings and a shared buried SiN/SiO$_2$ distributed
Bragg reflector (DBR) within one $24\times\SI{24}{\micrometre}$ emitting
footprint (Figure~\ref{fig:arch}). Under the stated aperture-conditioned
analysis, independently constructed Tidy3D and Lumerical models predict
relative focal separations below \SI{0.1}{\micrometre} after upper-layer
phase retargeting, with more than 99.9\% of each peak intensity available
at a common site. The corresponding Tidy3D focusing-efficiency metrics are
0.410 and 0.168; the qubit-channel positioning-disk crosstalk is approximately
$-18$~dB at \SI{5}{\micrometre} ion spacing. The broader contribution is a
design strategy that combines separately synthesized narrowband wavefronts
within a shared free-space interface. The calcium system supplies a concrete
numerical test of this strategy. The same hierarchy can be reapplied to
other wavelength pairs through appropriate material selection and
reoptimization of the local scatterers, phase profiles, and layer stack.

\section{Results and Discussion}

\subsection{Depth allocation as an apodization coordinate}

\begin{figure}[!t]
  \centering
  \includegraphics[width=0.94\linewidth]{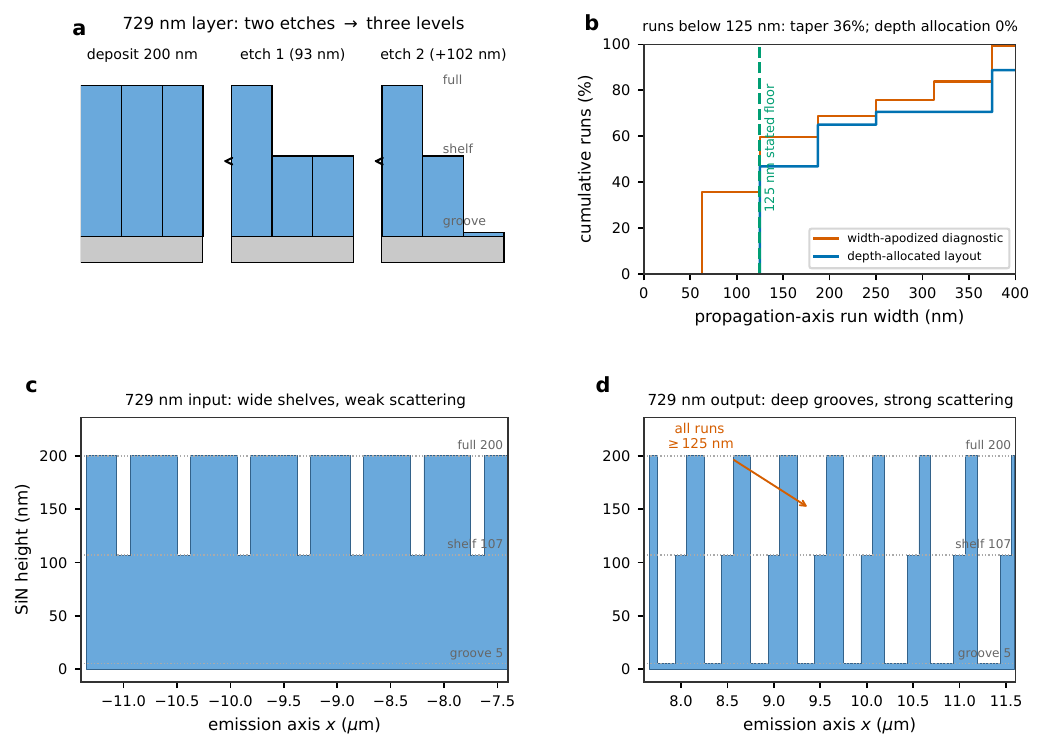}
  \caption{\textbf{Multi-level depth-allocation apodization.}
  Panels (a), (c), and (d) show the lower \SI{729}{\nano\meter} layer, whose full, shelf, and groove
  heights are 200, 107, and \SI{5}{\nano\meter}. (a) Two etches create these three SiN levels.
  (b) Propagation-axis run-width
  cumulative distribution: 36\% of runs in the width-apodized diagnostic fall below \SI{125}{\nano\meter},
  whereas none do so in the depth-allocated layout. (c) At the aperture input, wide shallow
  shelves provide weak scattering without narrow runs. (d) Toward the output, deep grooves
  provide stronger scattering while every checked propagation-axis line and space remains at
  least \SI{125}{\nano\meter}.}
  \label{fig:method}
\end{figure}

A focusing grating must simultaneously impose phase curvature and an amplitude envelope. Tooth
locations follow the confocal astigmatic phase of Suhara and Nishihara~\cite{Suhara1986}; the
local radiation-decay coefficient $\alpha(x)$ is then graded so that the guided power is emitted
over the available length. The familiar \aL$\approx1$ design regime fills the aperture efficiently,
but a width-only taper obtains weak scattering at the input by making one constituent of the
period progressively smaller. On the \SI{62.5}{\nano\meter} design raster, a representative
width-apodized diagnostic places 36\% of its propagation-axis runs below the adopted
\SI{125}{\nano\meter} floor (Figure~\ref{fig:method}).

Depth allocation uses the degrees of freedom already present in a two-etch process. Each
\SI{200}{\nano\meter} SiN film has three physical levels: full film, an intermediate shelf, and a
deep groove. Wide shallow shelves supply the weak input perturbation; deeper grooves are introduced
and widened toward the output as stronger scattering is required. The local period is included when
the feature floor is applied, and a row-wise run-length correction merges residual one-cell features
into the larger adjacent region. The resulting four sequential masks---two etches in each of two
layers---contain no sub-\SI{125}{\nano\meter} propagation-axis line or space under this stated
metric. This is discrete multi-level lithography, not continuous grayscale etching.

The final masks establish the fabrication level result: depth allocation implements a graded
radiation envelope while eliminating sub-threshold propagation-axis runs. The
width-taper reference in Figure~\ref{fig:method} is a geometric diagnostic rather than a
constraint-matched optimization: it shows how one conventional weak-scattering construction enters
the sub-threshold regime. Its full-wave optical response and the precise scope of the comparison are
reported in Supporting Information.

\subsection{Mask-level phase registration and two-color focus}

The device comprises a lower \SI{729}{\nano\meter} aperture and an upper
\SI{854}{\nano\meter} aperture separated by \SI{1}{\micrometre} of SiO$_2$. A nominal
\SI{0.50}{\micrometre} oxide standoff separates the lower film from a ten-pair quarter-wave DBR.
Here, phase registration means
adjusting one aperture's emitted wavefront so that its focal phase front reaches the same transverse
coordinate as the other color at the ion plane. It is implemented by retargeting the upper-layer
confocal phase coordinate by $-\SI{2.5}{\micrometre}$, from \SI{20.9}{\micrometre} to
\SI{18.4}{\micrometre}, and regenerating the upper tooth trajectories. The upper mask therefore
changes, but the \SI{24}{\micrometre} footprint, physical layer position, etch levels,
minimum-feature operation, effective index, and complete lower mask remain fixed. The
\SI{2.5}{\micrometre} value is a phase-target shift, not a physical interlayer overlay.

The final calculations are full-stack: both patterned
films and the DBR are present in every solve. Transmission of the lower-channel radiation through
the upper pattern, reciprocal loading of the active layer, and any resulting aberration or stray
radiation inside the modeled aperture therefore contribute to the reported focal fractions and
crosstalk. Beyond the aperture, the external model continues the excited film; active-port
reflection is therefore the routing metric reported here.

The layer order deliberately prioritizes the coherent qubit channel. The lower
\SI{729}{\nano\meter} grating lies closest to the reflector and receives the larger directional
enhancement, while the upper \SI{854}{\nano\meter} repump layer provides an independent wavefront
in the same footprint. The full-stack calculation includes the
wavefront distortion incurred as the 729-nm field traverses the upper patterned film.

\begin{figure}[!t]
  \centering
  \includegraphics[width=0.94\linewidth]{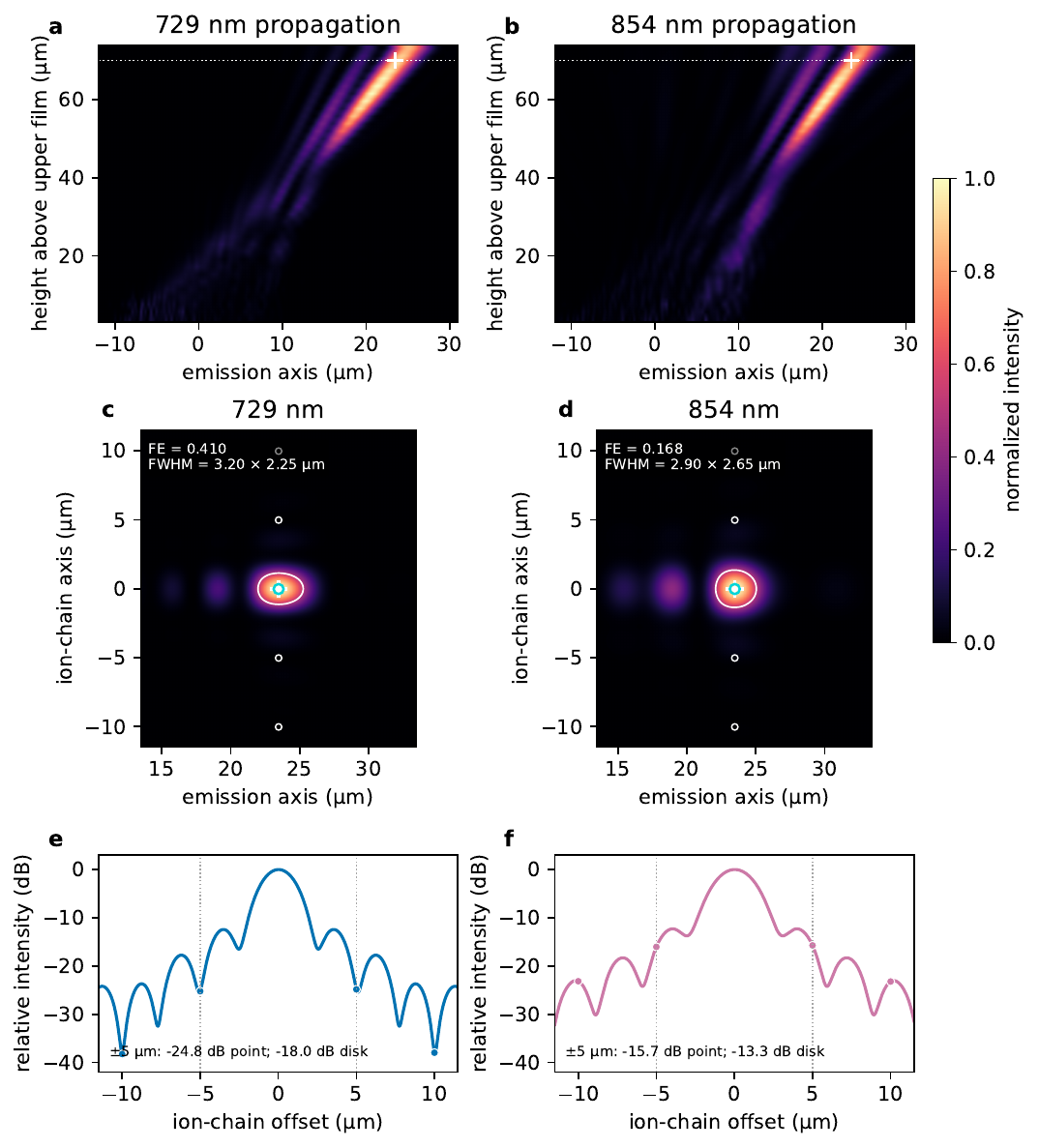}
  \caption{\textbf{Propagation and phase-registered focal fields.}
  (a,b) Emission-plane sections of the angular-spectrum-propagated vector intensity from the
  near-field plane to the ion plane; dashed lines mark the selected height. (c,d) Finest-mesh Tidy3D
  intensity maps at \SI{70}{\micrometre}; white contours show half maximum and cyan circles mark
  the common site. The individual maxima are unresolved on the \SI{50}{\nano\meter} reporting
  grid. (e,f) Ion-chain profiles through that site. The marked $\pm\SI{5}{\micrometre}$ samples
  give $-24.8$~dB at \SI{729}{\nano\meter} and $-15.7$~dB at \SI{854}{\nano\meter}; independent
  Lumerical values are $-24.26$ and $-15.94$~dB. The finite-position disk values are $-18.03$ and
  $-13.31$~dB in Tidy3D and $-17.70$ and $-13.26$~dB in Lumerical. FE denotes upward coupling
  multiplied by the fraction within a \SI{2}{\micrometre}-radius focal bucket.}
  \label{fig:focus}
\end{figure}

Figure~\ref{fig:focus} shows the final Tidy3D solution at $dx=dy=\SI{20}{\nano\meter}$ with
\SI{15}{\nano\meter} vertical refinement through the physical stack. At a height of
\SI{70}{\micrometre} measured from the upper-film surface, the two maxima fall on the same
\SI{50}{\nano\meter} reporting pixel; a
local quadratic fit gives a residual relative displacement of approximately
\SI{0.03}{\micrometre}. The common coordinate retains 100\% of both peak intensities to the quoted
precision. The \SI{729}{\nano\meter} and \SI{854}{\nano\meter} spots have emission-axis by
ion-chain FWHM values of $3.20\times\SI{2.25}{\micrometre}$ and
$2.90\times\SI{2.65}{\micrometre}$, respectively. Their chain-axis widths are close to the
diffraction scale imposed by the finite aperture; the larger emission-axis widths reflect the
oblique, apodized aperture and are reproduced by the independent solver.

The upward-coupling fractions are 0.711 and 0.381, and the fractions of the propagated field
within a \SI{2}{\micrometre}-radius focal bucket are 0.577 and 0.442. Their product gives focusing
efficiencies of 0.410 and 0.168 at \SI{729}{\nano\meter} and \SI{854}{\nano\meter}, respectively.
The corresponding directionality values are 0.840 and 0.541. Writing the upward-plus-downward
collected fraction as
$\eta_{\mathrm{ud}}=(P_{\uparrow}+P_{\downarrow})/P_{\mathrm{in}}
=\eta_{\uparrow}/D$ separates the two power factors:
$\eta_{\mathrm{ud}}=0.846$ and 0.704 at 729 and \SI{854}{\nano\meter}. The lower
focusing efficiency at 854 nm therefore reflects both incomplete extraction from the finite aperture and reduced
upward directionality; the focal fraction is comparable to that of the qubit channel. The large
directionality penalty has a clear spectral origin: the ten-pair quarter-wave DBR is centered at
\SI{700}{\nano\meter}. A finite-stack
transfer-matrix calculation at the operating angle gives power reflectances of 0.991 at
\SI{729}{\nano\meter} and 0.271 at \SI{854}{\nano\meter}; the latter wavelength lies outside the
$R\geq0.9$ band of approximately 619--\SI{773}{\nano\meter}. The full-device directionality also
depends on the grating and round-trip phase, so these values diagnose the spectral trade rather than
uniquely decompose it. Centering an otherwise identical quarter-wave stack near
\SI{802}{\nano\meter} gives reflectances of 0.9823 and 0.9822: the 729-nm reflectance decreases from
0.9909, and both reflection phases move. Because the gratings occupy different round-trip distances
from the reflector, this dual-line point requires joint reoptimization of the DBR standoff and both
aperture phases. The selected operating point prioritizes the coherent qubit channel while retaining
a co-registered repump field. The allowed
\SI{854}{\nano\meter} transition requires much less optical power than the quadrupole transition,
whereas neighboring-ion response is governed by the spatial crosstalk discussed below.

Along a \SI{5}{\micrometre}-spaced ion chain, the point crosstalk is $-24.8$~dB at
\SI{729}{\nano\meter} and $-15.7$~dB at \SI{854}{\nano\meter}. Taking the maximum intensity in a
\SI{0.5}{\micrometre}-radius disk around the neighboring site gives the more conservative values
$-18.0$ and $-13.3$~dB. The point value samples a local sidelobe minimum and varies with ion height
and aperture conditioning; the disk value is therefore used for headline comparisons. The
qubit-channel intensity profile provides spatial
discrimination relevant to coherent addressing. At \SI{854}{\nano\meter}, the point and disk
levels correspond to 2.7\% and 4.7\% of the target intensity. The upper channel is therefore
positioned for co-registered shared or global repumping; a site-selective variant would add
sidelobe suppression to the optimization objective.

\FloatBarrier

\subsection{Independent validation and numerical refinement}

The final GDS and physical layer stack were solved independently in Flexcompute Tidy3D and Ansys
Academic Research Lumerical FDTD 2026 R1.2. The finest Tidy3D calculation uses
$dx=dy=\SI{20}{\nano\meter}$, a \SI{15}{\nano\meter} vertical mesh override through the stack,
and subpixel interface treatment. Lumerical uses automatic lateral meshing (accuracy 2), a controlled
\SI{20}{\nano\meter} vertical stack override, and conformal interface treatment. Identical
complex-vector, distance-band-limited angular-spectrum propagation and focal
definitions are then applied to both solvers.

Table~\ref{tab:solver-main} summarizes the cross-solver result. Lumerical places the two maxima on
reporting pixels separated by \SI{0.05}{\micrometre}; local fits give
\SIrange{0.068}{0.073}{\micrometre}, and the shared site retains 99.92\% of each peak. Spot
dimensions agree within \SI{0.05}{\micrometre}, focusing efficiencies within 0.009 absolute, and
disk crosstalk within 0.33~dB. Within
each solver the colors are registered below \SI{0.1}{\micrometre}; their absolute common
emission-axis coordinates differ by approximately \SI{1.3}{\micrometre}. Relative two-color
registration is therefore a reproducible property of the fixed layout, while absolute mask-to-ion
placement remains a calibration parameter.

\begin{table}[htbp]
\centering
\caption{Independent full-stack validation. Entries are Tidy3D/Lumerical.}
\label{tab:solver-main}
\small
\begin{tabular}{lcc}
\toprule
Metric & \SI{729}{\nano\meter} & \SI{854}{\nano\meter} \\
\midrule
Directionality & 0.840/0.846 & 0.541/0.537 \\
Upward coupling & 0.711/0.705 & 0.381/0.367 \\
Focusing efficiency & 0.410/0.402 & 0.168/0.162 \\
FWHM, emission $\times$ chain (\si{\micrometre}) & $3.20\times2.25$/$3.25\times2.20$ & $2.90\times2.65$/$2.95\times2.70$ \\
Point crosstalk at \SI{5}{\micrometre} (dB) & $-24.84$/$-24.26$ & $-15.73$/$-15.94$ \\
Disk crosstalk at \SI{5}{\micrometre} (dB) & $-18.03$/$-17.70$ & $-13.31$/$-13.26$ \\
\bottomrule
\end{tabular}
\end{table}

Lateral refinement provides the decisive registration check. In Tidy3D, the peak separation falls
from \SI{0.50}{\micrometre} at \SI{33.3}{\nano\meter} lateral resolution to
\SI{0.15}{\micrometre} at \SI{25}{\nano\meter} and zero reporting pixels at
\SI{20}{\nano\meter}; the shared-site retention rises from 98.3\% to 99.8\% and 100\%. Optical
efficiencies vary smoothly over the same sequence. This controlled Tidy3D refinement, followed by
an independently built Lumerical model with different meshing and interface treatment, supports
subwavelength co-registration rather than a single-grid coincidence. Lumerical was not assigned an
in-plane grid spacing because its lateral mesh remains automatic. Detailed tables, centroid and
local-fit separations, and the solver conventions are
given in Supporting Information.

The nominal subpixel separation is more precise than the operational requirement. Translating one
final focal map relative to the other while reselecting the shared site shows that each color remains
within 2\% of its own maximum for relative displacements up to \SI{0.50}{\micrometre} along the
emission axis and \SI{0.40}{\micrometre} along the ion chain. This field-derived margin quantifies
registration tolerance without assigning it to a particular overlay, thickness, or material-index
distribution.

\subsection{Ion-height operating range and electrode-footprint placement}

The common focus is robust over a practical range of ion heights. Repropagating the same finest-mesh
near fields from 60 to \SI{85}{\micrometre}, measured from the upper-film surface, gives 99.0\%
shared-site retention at \SI{60}{\micrometre} and more than 99\% from 62.5 to
\SI{75}{\micrometre}, with a maximum near \SI{70}{\micrometre}. That plane also
minimizes the \SI{729}{\nano\meter} neighbor intensity: the point crosstalk improves from
$-19.5$~dB at \SI{65}{\micrometre} to $-24.8$~dB at \SI{70}{\micrometre}, then relaxes to
$-20.5$~dB by \SI{75}{\micrometre}. Within that interval, \SI{70}{\micrometre} gives the best combination
of relative registration and qubit-channel neighbor suppression.

\begin{figure}[htbp]
  \centering
  \includegraphics[width=\linewidth]{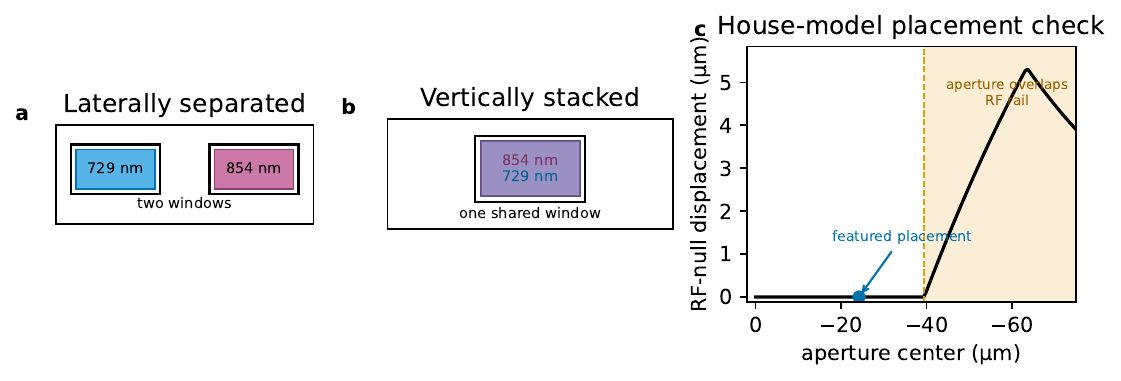}
  \caption{\textbf{Electrode-window economy and placement.}
  (a) A conventional layout with one opening per laterally separated emitter. (b) Vertically
  stacked apertures overlap the two emitting footprints in one window. (c) Scaled placement in the
  gapless House conductor-plane model. At the corresponding optical plane, \SI{72.13}{\micrometre}
  above the upper film, the common coordinate is laterally aligned with the symmetry-plane RF null
  by centering the \SI{24}{\micrometre} aperture at $-\SI{24.225}{\micrometre}$, leaving
  \SI{15.275}{\micrometre} to the inner rail boundary. The zero shift follows from replacing the
  aperture by the same grounded boundary. The panel presents footprint placement rather than a
  dielectric-window field calculation.}
  \label{fig:trap}
\end{figure}

Vertical stacking also changes the electrode-level geometry. In the conventional reference layout
shown in Figure~\ref{fig:trap}, each laterally separated grating uses its own window; the bilayer
overlaps their emitting footprints and uses one. The unperturbed House null is
\SI{70.13}{\micrometre} above its conductor plane. If that plane coincides with the outer surface
of the nominal \SI{2}{\micrometre} overclad, the corresponding optical height is
\SI{72.13}{\micrometre} above the upper film. Direct repropagation to that plane places the common
optical coordinate \SI{24.225}{\micrometre} from the grating center, with 99.93\% shared-site
retention and $-23.58$~dB 729-nm point crosstalk. Centering the proposed
\SI{24}{\micrometre} opening at $-\SI{24.225}{\micrometre}$ therefore aligns that coordinate with
the symmetry line of the analytic, gapless surface-electrode model of House~\cite{House2008}. The
nearest aperture edge is then \SI{15.275}{\micrometre} from the inner RF-rail boundary at
$-\SI{51.5}{\micrometre}$; first contact would occur only when the aperture center reaches
$-\SI{39.5}{\micrometre}$. Because the aperture
is represented by the same grounded boundary as the central electrode, the calculated null shift
is unchanged by construction. The calculation establishes scaled footprint placement and RF-rail
clearance under the analytic House boundary. A realized opening requires a device-level trap model
including dielectric permittivity and charge, finite metal thickness and gaps, secular confinement,
and transparent shielding.

The system-level benefit is the reuse of the ion-facing emission footprint
while retaining separate optical inputs and wavelength-specific phase
profiles. Relative to the laterally separated reference geometry, the
demonstrated bilayer combines two emitting regions within one proposed
electrode opening. Additional spectral channels would require further
layers, routing, and joint electromagnetic optimization: the architecture
trades lateral aperture area for vertical process complexity. The local
scatterer search, analytic phase synthesis, and discrete depth allocation
provide a design procedure for making that trade while enforcing an explicit
longitudinal feature constraint.

This procedure is not specific to the calcium transitions. A concrete
extension is the \SI{493}{\nano\meter} cooling and
\SI{650}{\nano\meter} repumping pair of
$\mathrm{Ba}^{+}$~\cite{Klawson2026}. Such an extension would require
wavelength-specific material data and reoptimization of guided-mode
confinement, grating scattering and phase, reflector response, and interlayer
coupling. The target operation also sets the polarization and power
requirements. Because the fabrication floor is an absolute length, changing
wavelength cannot in general be achieved by uniformly rescaling the present
mask. More broadly, chip-based multicolor excitation of neutral atoms and
compact spectroscopy are candidate applications when independently shaped
fields must share restricted optical access. Their required numerical
apertures, wavelength ranges, and operating powers must be evaluated in new
device designs; the present numerical demonstration concerns the calcium
wavelength pair.

For the quadrupole transition, the transition-specific Rabi frequency depends on the field-gradient
tensor and quantization-axis geometry,
$\Omega\propto\sum_{ij}\langle D,m'|Q_{ij}|S,m\rangle\,\partial_iE_j$
~\cite{James1998,Verde2023}. Reprocessing the exported vector field gives a normalized,
orientation-independent quadrupolar-gradient proxy
$G=\|\operatorname{sym}_0(\nabla\mathbf E)\|_F^2$. At the \SI{5}{\micrometre} neighbors,
$G/G_0$ is $-23.0$~dB at \SI{729}{\nano\meter} and $-15.6$~dB at
\SI{854}{\nano\meter}, close to the corresponding intensity ratios. This tensor norm verifies that
the spatial discrimination persists in a vector-gradient description. A selected Zeeman line and
magnetic-field direction contract this tensor with $Q_{ij}$; the pulse sequence and spectator
detuning then determine the corresponding gate error.

\section{Methods}

\textbf{Geometry and materials.}
The physical stack, from bottom to top, is SiO$_2$ substrate; ten SiN/SiO$_2$ DBR pairs with
\SI{86.72}{\nano\meter}/\SI{120.44}{\nano\meter} thicknesses; a nominal
\SI{0.50}{\micrometre} SiO$_2$ standoff; the \SI{200}{\nano\meter} lower SiN grating;
\SI{1.00}{\micrometre} of interlayer SiO$_2$; the \SI{200}{\nano\meter} upper SiN grating; and
\SI{2.00}{\micrometre} of upper cladding. The lower three-level heights are
5/107/\SI{200}{\nano\meter}; the upper heights are 5/123/\SI{200}{\nano\meter}. Refractive
indices are $n_{\mathrm{SiN}}=2.018$ and $n_{\mathrm{SiO_2}}=1.453$ at both discrete operating
wavelengths. The nominal model uses lossless, nondispersive materials and vertical sidewalls; the
two-solver comparison therefore isolates geometry and discretization under common constitutive
assumptions. Film-specific dispersion, absorption, roughness, and etch taper enter during process
transfer from measured material and cross-sectional data.
The common focal coordinate corresponds to an \SI{18.5}{\degree} emission angle in air.

\textbf{Hierarchical adjoint-assisted construction and phase registration.}
A differentiable three-level unit cell was optimized in a single-film, mirror-free FDTDX
model~\cite{Mahlau2026} over four bounded
parameters: deep-groove width, shallow-shelf width, and the two etch depths. The objective rewards
upward coupling and directionality. The cell is tiled along an astigmatic confocal phase whose
local derivative sets the chirped grating period, and it is modulated by the depth-allocation rule
$s(x)=s_{\min}+(1-s_{\min})t(x)^p$, where $t$ runs from the aperture input to output. The deep
groove is absent at small $s$ and is introduced as $s$ increases. A local-period-aware two-cell
floor and row-wise run-length correction enforce the stated \SI{125}{\nano\meter} propagation-axis
line and space. The full $24\times\SI{24}{\micrometre}$ bilayer aperture consequently uses a
structured hierarchical design in place of a free-voxel parameterization. The hierarchy assigns gradient optimization to the compact
local-scattering problem, analytic synthesis to the known confocal phase, and deterministic depth
allocation to the foundry rule. This reduces the search dimension and keeps the phase law and
minimum-feature transformation directly traceable to physical design variables.

The final registration step is a low-dimensional physical refinement rather than a new topology
optimization. The upper-layer confocal target parameter was screened in Tidy3D while the lower
layer and all stack and fabrication parameters were held fixed. Candidate shifts were screened at
\SI{40}{\nano\meter} lateral resolution. The leading neighborhood was narrowed through
two-candidate calculations at \SI{33.3}{\nano\meter} and \SI{25}{\nano\meter}; because the ordering
changed with refinement, the selected $-\SI{2.5}{\micrometre}$ phase retargeting was evaluated at
both \SI{33.3}{\nano\meter} and \SI{20}{\nano\meter}. Both wavelengths were rerun for every candidate with both patterned films
present, so transmission through and parasitic loading by the other grating were included. Only the
excited film was continued into the external lead, making active-port reflection the port-level
quantity of this aperture calculation. The fixed final GDS was then rebuilt
independently in Lumerical.

\textbf{Electromagnetic validation.}
The final predictions use Flexcompute Tidy3D v2.11.2 and Ansys Academic Research Lumerical FDTD
2026 R1.2~\cite{Tidy3D,LumericalFDTD}. Each channel is excited by the fundamental transverse-electric
mode of its active slab and simulated separately at its operating wavelength. Tidy3D uses subpixel
interface averaging, perfectly matched layers, and the mesh sequence stated above. Lumerical uses
automatic lateral meshing (accuracy 2), conformal interface treatment, and a \SI{20}{\nano\meter}
vertical stack override. Upward and downward Poynting fluxes are recorded
above and below the device, and input power is obtained from a guided-mode monitor. Directionality
is $D=P_{\uparrow}/(P_{\uparrow}+P_{\downarrow})$, and upward coupling is
$\eta_{\uparrow}=P_{\uparrow}/P_{\mathrm{in}}$.

The complex vector near field is sampled \SI{0.5}{\micrometre} above the upper cladding and
propagated \SI{67.5}{\micrometre} using the band-limited angular-spectrum
method~\cite{Matsushima2009}. The target is therefore \SI{70}{\micrometre} above the upper-film
surface and \SI{68}{\micrometre} above the nominal outer cladding surface.
The same super-Gaussian aperture conditioner, with field exponent 16 (order 8 in intensity), and the same
$|E_x|^2+|E_y|^2+|E_z|^2$ definition are applied to both solvers. Focusing efficiency is
$\mathrm{FE}=\eta_{\uparrow}f_{2\mu\mathrm{m}}$, where $f_{2\mu\mathrm{m}}$ is the fraction of the
propagated intensity inside a \SI{2}{\micrometre}-radius disk centered on the focal maximum.
Point crosstalk is the intensity at $\pm\SI{5}{\micrometre}$ along the ion-chain axis relative to
the target site; disk crosstalk is the worst value in a \SI{0.5}{\micrometre}-radius neighborhood of
that ion coordinate. FWHM is measured on the connected half-maximum component containing the
selected peak.
The angular-spectrum implementation was checked independently against analytic Gaussian-beam
propagation; the propagated waist and on-axis intensity agree within 0.5\% over the reported
distance. 

\textbf{Mask and trap calculations.}
The four sequential etch masks were checked row by row and screened morphologically in vector
form. The \SI{125}{\nano\meter} statement refers specifically to propagation-axis line and space;
the staircase representation retains smaller boundary jogs. Foundry release then applies the target
process rule deck, bias, and corner treatment. The trap-placement curve uses the analytic, gapless surface-electrode boundary
solution of House~\cite{House2008}; it evaluates radial RF-null displacement as the aperture center
is moved toward the rail.

\section{Conclusion}

This work establishes a scalable design hierarchy for multicolor quantum-photonic emitters that decouples addressing wavelength count from chip-surface optical real estate. By mapping independent spectral channels to vertically stacked planes, confining differentiable optimization to the nonlinear local scattering of a three-level unit cell, and realizing the aperture radiation envelope through deterministic depth allocation across discrete etches, the architecture enforces an explicit \SI{125}{\nano\meter} propagation-axis critical-dimension floor without grayscale patterning. Subsequent full-stack phase retargeting locks the emitted wavefronts to a common microscopic coordinate, converting multi-color delivery from a repeated surface-area penalty into an interpretable vertical wavefront-allocation problem.

Applied to \Ca{}, the monolithic bilayer device focuses the \SI{729.4}{\nano\meter} qubit and \SI{854.2}{\nano\meter} repump transitions from a single $24\times\SI{24}{\micrometre}$ aperture onto a shared target \SI{70}{\micrometre} above the chip. Independent full-stack evaluations using mesh-converged Tidy3D and independently built Lumerical FDTD models reproduce sub-\SI{0.1}{\micrometre} focal co-registration, $>99.9\%$ shared-site peak intensity retention, and focusing efficiencies of 0.410 and 0.168. At \SI{5}{\micrometre} ion spacing, the qubit channel achieves $-24.8$~dB point crosstalk, remains bounded below $-18.0$~dB across a conservative \SI{0.5}{\micrometre}-radius positioning disk, and yields $-23.0$~dB quadrupolar field-gradient suppression. Across an experimental vertical alignment window of \SIrange{67.5}{72.5}{\micrometre}, shared-site retention exceeds 99.85\% while point crosstalk remains below $-22.8$~dB. Crucially, under the analytic House conductor-plane model, the unified aperture provides \SI{15.3}{\micrometre} clearance to the nearest radio-frequency rail with zero RF-null displacement, replacing two lateral openings with one modeled window and mitigating light-induced charging of exposed dielectrics.

By uniting mask-locked co-alignment, transition-specific wavefront tailoring, and strict foundry design rules within a solitary surface footprint, this approach removes a major integration bottleneck in surface-electrode traps, multizone QCCD processors, and dense neutral-atom tweezer arrays. Future extensions incorporating dual-band distributed Bragg reflectors and process-specific robust optimization offer direct routes to enhanced repump collection and high fabricated yields. The demonstrated framework provides a transferable, foundry-ready foundation for co-registering complementary optical transitions—such as the \SI{493}{\nano\meter}/\SI{650}{\nano\meter} lines in $\mathrm{Ba}^{+}$ or dual-wavelength Rydberg trapping fields—wherever multi-chromatic control must address microscopic quantum systems through constrained device interfaces.

\section*{Supporting Information}
Detailed design parameters and layer definitions; phase-registration and mesh-convergence tables;
independent-solver metrics; propagation, conditioning, and registration-margin checks; reflector
trade analysis; prior-art metric comparison; ion-height scan; crosstalk definitions; mask and
comparator analyses; design-stage FDTDX resolution; and the House-model placement calculation (PDF).
The Supporting Information also includes a matched zero-residual-layer sensitivity calculation.

\section*{Funding}
Centre for Quantum Technologies (CQT), National University of Singapore --- Young Researcher
Career Development Grant no. 26-YRCDG-GY awarded to G.~Y.

\section*{Acknowledgments}
The author thanks Manas Mukherjee and Yanyan Zhou for helpful discussions. This work used Vanda 
and Atlas high-performance computing resources of the National University of Singapore. OpenAI Codex was used for language editing, Latex drafting. Claude Code (Anthropic) was used for code development and troubleshooting; all computational simulations and runs were independently executed and verified by the author. No generative-AI imagery is used;
the three-dimensional device view is a physically based rendering of the final numerical height maps.

\section*{Notes}
The author declares no competing financial interest.

\section*{Data Availability Statement}
The final device, fabrication layout, Tidy3D and Lumerical field data, analysis code, and scripts
used to reproduce every reported figure and numerical table are assembled in a versioned reviewer
archive. The archive will be deposited in a public repository, and its persistent identifier will
be added to the version of record.

\bibliographystyle{achemso}
\bibliography{references}

\clearpage
\section*{For Table of Contents Only}
\begin{center}
  \includegraphics[width=3.25in,height=1.75in,keepaspectratio]{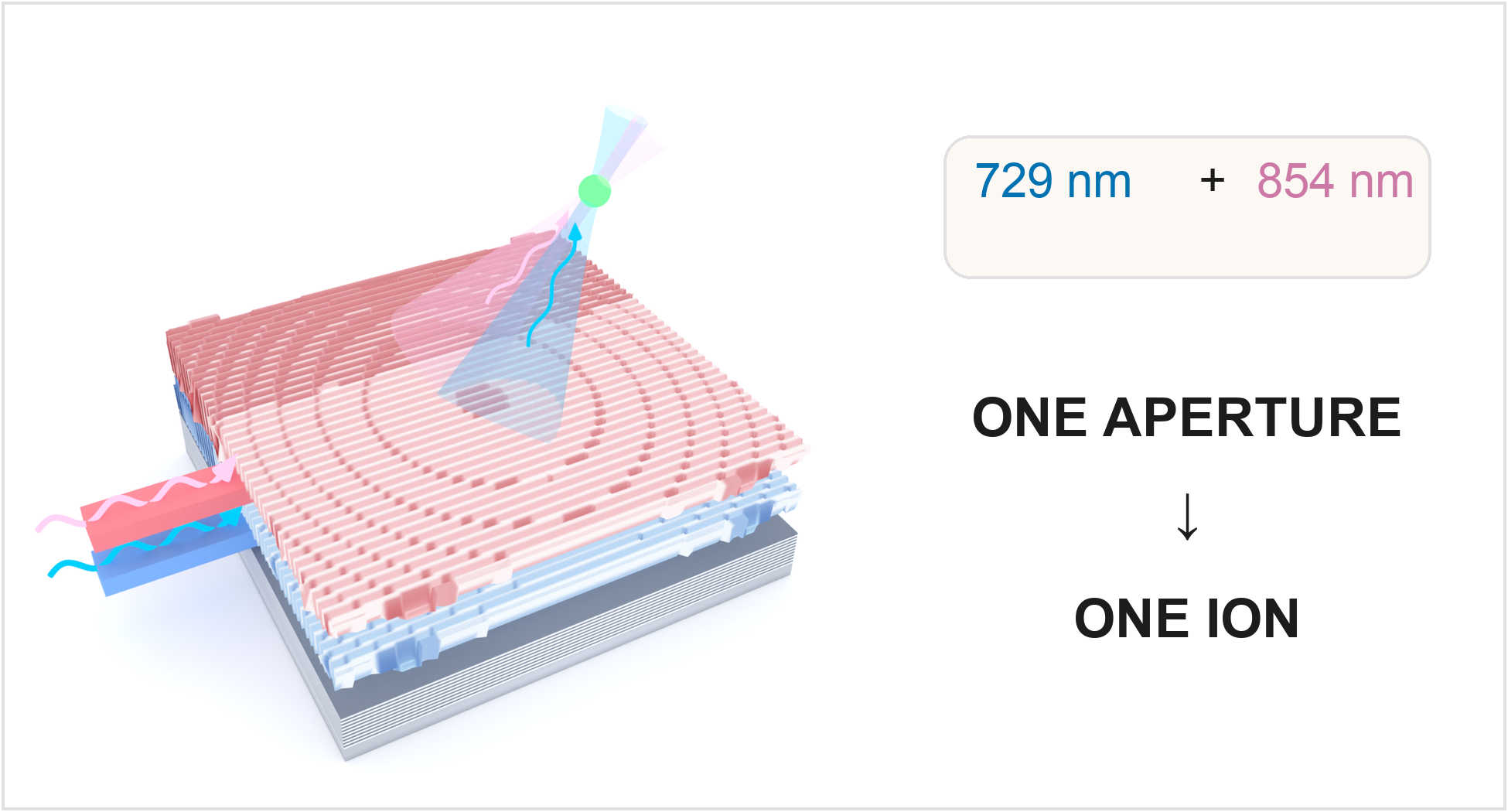}
\end{center}

\end{document}


\maketitle

This Supporting Information reports the numerical definitions and refinement evidence for the
fixed, phase-registered device. Final quantitative claims use two independent physical-stack
solvers: Flexcompute Tidy3D v2.11.2 and Ansys Academic Research Lumerical FDTD 2026 R1.2.
FDTDX is identified separately wherever it records the differentiable design-stage model.

\section{Device and layer definition}

The two grating planes occupy the same $24\times\SI{24}{\micrometre}$ lateral region within a
shared optical footprint intended for one electrode opening. The lower SiN film carries the
\SI{729.4}{\nano\meter} radiator and the upper
film carries the \SI{854.2}{\nano\meter} radiator. Both are independently fed along the positive
emission axis.
Table~\ref{tab:stack} gives the nominal physical stack used in the final Tidy3D and Lumerical
calculations.

\begin{table}[htbp]
\centering
\caption{Nominal physical stack, from bottom to top.}
\label{tab:stack}
\small
\begin{tabular}{lll}
\toprule
Element & Material & Thickness or level height \\
\midrule
Substrate & SiO$_2$ & semi-infinite in the physical device \\
Buried reflector & 10 $\times$ (SiN/SiO$_2$) & 86.72/\SI{120.44}{\nano\meter} \\
Reflector standoff & SiO$_2$ & \SI{0.50}{\micrometre} \\
Lower three-level aperture & SiN & 5/107/\SI{200}{\nano\meter} \\
Interlayer & SiO$_2$ & \SI{1.00}{\micrometre} \\
Upper three-level aperture & SiN & 5/123/\SI{200}{\nano\meter} \\
Upper cladding & SiO$_2$ & \SI{2.00}{\micrometre} \\
\bottomrule
\end{tabular}
\end{table}

The final refractive indices are $n_{\mathrm{SiN}}=2.018$ and
$n_{\mathrm{SiO_2}}=1.453$. The operating-point calculations treat both as real and constant at
the two discrete wavelengths. The production GDS contains separate layer/datatype pairs for the
two etches in each film. Its imported polygon bounds are $(-12,-12)$ to
$(12,12)~\si{\micrometre}$, and the exact layer filter was verified before every final rebuild.

The height maps use propagation coordinate $x$, ion-chain coordinate $y$, and aperture length
$L=\SI{24}{\micrometre}$. Their separable astigmatic path function and normalized phase are
\begin{align}
\ell(x,y)&=\sqrt{(x-x_0)^2+h_x^2}+\sqrt{y^2+h_y^2},\\
q(x,y)&=\frac{1}{2\pi}\operatorname{mod}\!\left[
\frac{2\pi}{\lambda}\{n_{\mathrm{eff}}x+\ell(x,y)\},2\pi\right].
\end{align}
Here $x_0$ translates the propagation-axis phase target, while $h_x$ and $h_y$ set the independent
phase curvatures along the emission and ion-chain axes. They are synthesis lengths rather than the
physical ion height. Differentiating the phase gives the local period
\begin{equation}
\Lambda(x)=\frac{\lambda}{\left|n_{\mathrm{eff}}+
(x-x_0)/\sqrt{(x-x_0)^2+h_x^2}\right|}.
\end{equation}

The longitudinal coordinate $t=\operatorname{clip}[(x+L/2)/L,0,1]$ defines the allocation weight
$s(x)=s_{\min}+(1-s_{\min})t^p$. With the fractional feature floor
$r(x)=m_{\min}/\Lambda(x)$, the shelf and deep fractions of a local period are
\begin{align}
f_s(x)&=\max(f_{s0},r),\\
f_d(x)&=
\begin{cases}
0,&d\,s(x)<r,\\
\min[d\,s(x),1-r-f_s(x)],&d\,s(x)\ge r.
\end{cases}
\end{align}
The height is the deep level for $q<f_d$, the shelf level for
$f_d\le q<f_d+f_s$, and the full film otherwise. A final row-wise repair merges any
propagation-axis run shorter than two \SI{62.5}{\nano\meter} design pixels into the larger adjacent
region. Table~\ref{tab:design-parameters} lists every free parameter in these equations. The
effective indices are synthesis parameters, not fitted material indices.

\begin{table}[htbp]
\centering
\caption{Final height-map synthesis parameters. Shelf and deep-width entries are fractions of the
local period before the feature-floor operation.}
\label{tab:design-parameters}
\small
\resizebox{\linewidth}{!}{%
\begin{tabular}{lccccccccc}
\toprule
Channel & $n_{\mathrm{eff}}$ & $x_0$ (\si{\micrometre}) & $h_x/h_y$ (\si{\micrometre}) &
Period range (\si{\nano\meter}) & $m_{\min}$ (nm) & $s_{\min}$ & $p$ & $f_{s0}$ & $d$ \\
\midrule
729 nm & 1.58 & 20.9 & 118/90 & 485--556 & 130 & 0.20 & 2.5 & 0.22 & 0.50 \\
854 nm & 1.53 & 18.4 & 118/90 & 579--667 & 130 & 0.35 & 2.5 & 0.22 & 0.39 \\
\bottomrule
\end{tabular}}
\end{table}

The \SI{5}{\nano\meter} lowest level is the nominal residual height of the nearly fully etched
state, not a specified etch-stop layer. It is thinner than a full cell in both final meshes and is
represented by Tidy3D subpixel averaging and Lumerical conformal treatment. The reported
cross-solver result therefore corresponds to this explicitly defined nominal interface geometry.
As a direct representation check, both nominal \SI{5}{\nano\meter} SiN levels were replaced by
SiO$_2$ and both wavelengths were rerun at \SI{33.3}{\nano\meter} lateral resolution with the
\SI{15}{\nano\meter} stack refinement. Relative to the matched residual-layer calculation, the
peak separation changed from 0.50 to \SI{0.55}{\micrometre}, shared-site retention from 98.28\%
to 98.09\%, and FE from 0.4158 to 0.4146 at \SI{729}{\nano\meter} and from 0.1616 to 0.1563 at
\SI{854}{\nano\meter}; both FWHM pairs were unchanged. The residual layer therefore does not
control the focal registration or the reported optical performance.

\section{Full-stack interaction and active-port response}

Both patterned SiN films are present in every final Tidy3D and Lumerical solve. Consequently, the
lower-channel field propagates through the patterned upper film, and the reported upward power,
focal fraction, sidelobes, and crosstalk include the resulting phase perturbation and scattering.
The two colors are solved separately because they are driven by independent lasers; no mutual
optical phase between \SI{729}{\nano\meter} and \SI{854}{\nano\meter} is assumed or required.

The simulated input region continues the excited SiN film outside the patterned
$24\times\SI{24}{\micrometre}$ aperture. This boundary condition isolates the aperture response
from a routing-specific feed network. At the active input plane, summing the four recorded guided modes gives reflected-to-forward power
ratios of $2.4\times10^{-3}$ ($-26.2$~dB) at \SI{729}{\nano\meter} and
$1.3\times10^{-3}$ ($-28.9$~dB) at \SI{854}{\nano\meter}.

\section{Shared-reflector spectral trade}

The reflector thicknesses, \SI{86.72}{\nano\meter} of SiN and
\SI{120.44}{\nano\meter} of SiO$_2$, are exactly quarter-wave at
\SI{700}{\nano\meter} for the nominal indices. The final common coordinate corresponds to an
air-side angle of \SI{18.54}{\degree}; Snell's law gives \SI{12.64}{\degree} in the oxide above the
reflector. A ten-pair transfer-matrix calculation for s polarization at this internal angle gives
$R_{729}=0.9909$ and $R_{854}=0.2710$. The $R\geq0.9$ interval of this finite stack is approximately
619--\SI{773}{\nano\meter}, so the repump wavelength lies outside the high-reflectance band.

\begin{figure}[htbp]
\centering
\includegraphics[width=0.98\linewidth]{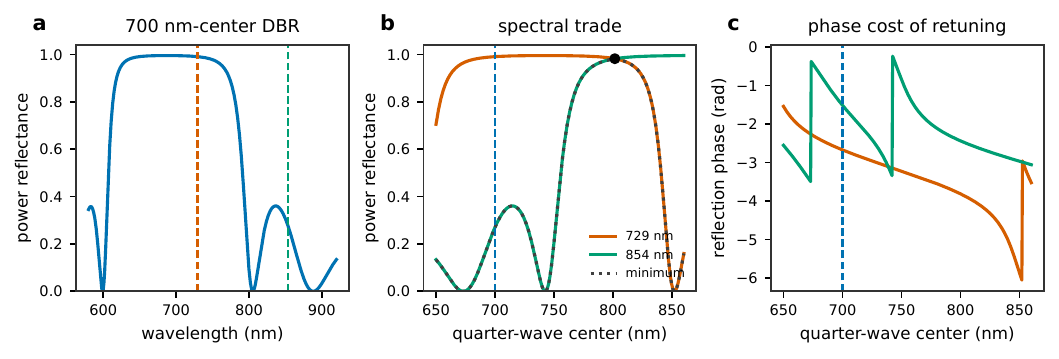}
\caption{\textbf{Finite-stack reflector trade.} (a) Power reflectance of the selected
\SI{700}{\nano\meter}-center, ten-pair stack at the operating internal angle. (b) Reflectance at
the two operating wavelengths as the quarter-wave center is varied. The black curve is the smaller
of the two reflectances. (c) Reflection phase over the same sweep, showing why spectral retuning
requires a new standoff and aperture-phase optimization.}
\label{fig:dbr-trade}
\end{figure}

Figure~\ref{fig:dbr-trade} separates this spectral fact from the full-device result. Retuning an
otherwise identical quarter-wave stack gives a maximin point near \SI{801.5}{\nano\meter}, where
the one-dimensional reflectances are 0.9823 and 0.9822. Relative to the selected stack,
the 729-nm reflectance decreases from 0.9909 while the 854-nm reflectance increases from 0.2710.
The same change also shifts the reflection
phase at both wavelengths. Because the lower and upper gratings have different round-trip distances
to the reflector, a dual-line design must optimize the DBR and both standoffs together with the two
aperture phases. The one-dimensional calculation identifies a high-value design direction; it does
not predict the directionality of a patterned bilayer by itself.

\section{Phase-registration study}

The initial physical-stack calculation showed that the two independently designed apertures had
similar focal shapes but a residual lateral registration error. A targeted study varied only the
upper-layer confocal phase target, leaving the lower mask, etch levels, reflector, aperture,
minimum-feature correction, source, and propagation pipeline unchanged. The selected design shifts
the upper phase target from \SI{20.9}{\micrometre} to \SI{18.4}{\micrometre}. Because the phase
map is regenerated, this operation changes the upper grating mask while physical interlayer overlay
remains zero.

Candidate upper-layer phase shifts were evaluated using progressively refined lateral grids from
\SI{40}{\nano\meter} to \SI{20}{\nano\meter}. The resulting
$-\SI{2.5}{\micrometre}$ phase retargeting was evaluated at \SI{33.3}{\nano\meter} and on the final
\SI{20}{\nano\meter} mesh, where it provides sub-pixel focal registration.
Table~\ref{tab:phaseconv} reports that candidate, with the reference design showing the gain from
phase registration.

\begin{table}[htbp]
\centering
\caption{Tidy3D phase-registration convergence at \SI{70}{\micrometre}. All runs use a
\SI{15}{\nano\meter} vertical stack mesh and subpixel treatment. FE values refer to the
individual focal maxima.}
\label{tab:phaseconv}
\small
\begin{tabular}{lccccc}
\toprule
Design / lateral mesh & Peak separation (\si{\micrometre}) & Shared site (\%) & FE$_{729}$ & FE$_{854}$ & $D_{729}/D_{854}$ \\
\midrule
Unregistered / 20 nm & 1.50 & 85.24 & 0.398 & 0.155 & 0.822/0.501 \\
Registered / 33.3 nm & 0.50 & 98.28 & 0.416 & 0.162 & 0.846/0.529 \\
Registered / 25 nm & 0.15 & 99.81 & 0.413 & 0.166 & 0.842/0.537 \\
Registered / 20 nm & 0.00 & 100.00 & 0.410 & 0.168 & 0.840/0.541 \\
\bottomrule
\end{tabular}
\end{table}

The peak maps were tabulated on a common \SI{50}{\nano\meter} reporting grid. At the finest mesh,
both Tidy3D maxima occupy $(23.475,-0.025)~\si{\micrometre}$. A local least-squares quadratic fit
to log intensity, repeated over $3\times3$, $5\times5$, $7\times7$, and $9\times9$ neighborhoods,
gives separations of 0.0303, 0.0301, 0.0297, and \SI{0.0293}{\micrometre}. These subpixel fits are
diagnostics rather than independent resolution; the primary Tidy3D result is coincidence within one
\SI{50}{\nano\meter} reporting pixel. The intensity centroid separation is
\SI{0.0148}{\micrometre}.

\begin{figure}[htbp]
\centering
\includegraphics[width=0.58\linewidth]{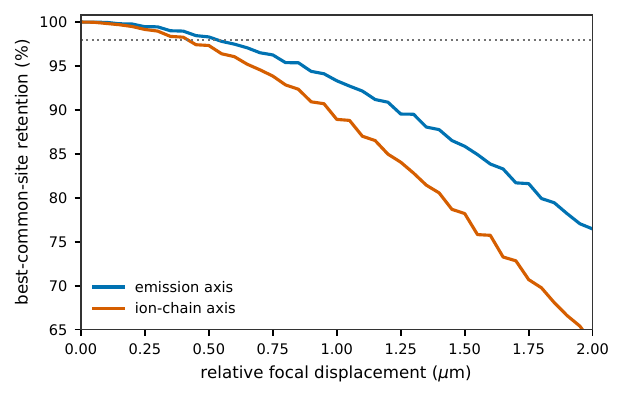}
\caption{\textbf{Operational registration margin.} Best-common-site retention after translating
one final focal field relative to the other along either transverse axis.}
\label{fig:registration-margin}
\end{figure}

The final 854-nm focal map was translated relative to the
729-nm map and the best shared site was reselected. Figure~\ref{fig:registration-margin} shows that
shared-site retention remains at least 98\% for relative displacements up to
\SI{0.50}{\micrometre} along the emission axis and \SI{0.40}{\micrometre} along the ion chain; the
99\% limits are \SI{0.35}{\micrometre} and \SI{0.25}{\micrometre}, respectively. A scalar
phase-gradient estimate gives
$\partial x_f/\partial n_{\mathrm{eff}}=h/\cos^3\theta\simeq\SI{82}{\micrometre}$ per unit
effective index at $h=\SI{70}{\micrometre}$ and $\theta=18.54^{\circ}$. The
\SI{0.50}{\micrometre} emission-axis margin therefore corresponds to an absolute differential
synthesis-index shift of approximately $6.1\times10^{-3}$. This is a useful phase-sensitivity scale;
a process tolerance requires the dispersion and thickness response of the selected films.

\section{Independent-solver validation}

The final GDS and nominal physical stack were independently built in Lumerical. The Lumerical solve
uses automatic lateral meshing (accuracy 2), a \SI{20}{\nano\meter} vertical refinement through the
DBR and active films, and conformal interface treatment. The near-field plane, distance-band-limited
angular-spectrum propagation, conditioning window,
reporting grid, focal bucket, FWHM, and crosstalk definitions are identical to the Tidy3D analysis.
Table~\ref{tab:solver} compares the final results.

\begin{table}[htbp]
\centering
\caption{Final physical-stack comparison. The first number is Tidy3D
($dx=dy=\SI{20}{\nano\meter}$, \SI{15}{\nano\meter} stack refinement); the second is Lumerical
(automatic lateral mesh and \SI{20}{\nano\meter} vertical refinement).}
\label{tab:solver}
\small
\begin{tabular}{lcc}
\toprule
Metric & \SI{729}{\nano\meter}: Tidy3D / Lumerical & \SI{854}{\nano\meter}: Tidy3D / Lumerical \\
\midrule
Directionality & 0.840 / 0.846 & 0.541 / 0.537 \\
Upward coupling & 0.711 / 0.705 & 0.381 / 0.367 \\
Focal fraction (\SI{2}{\micrometre} radius) & 0.577 / 0.570 & 0.442 / 0.441 \\
Focusing efficiency & 0.410 / 0.402 & 0.168 / 0.162 \\
FWHM, emission $\times$ chain (\si{\micrometre}) & $3.20\times2.25$ / $3.25\times2.20$ & $2.90\times2.65$ / $2.95\times2.70$ \\
Point crosstalk at \SI{5}{\micrometre} (dB) & $-24.84$ / $-24.26$ & $-15.73$ / $-15.94$ \\
Disk crosstalk at \SI{5}{\micrometre} (dB) & $-18.03$ / $-17.70$ & $-13.31$ / $-13.26$ \\
\bottomrule
\end{tabular}
\end{table}

The Tidy3D/Lumerical peak separations are 0.00/\SI{0.05}{\micrometre}; the corresponding
best-common-site retentions are 100.00/99.92\%. Lumerical centroid separation is
\SI{0.0499}{\micrometre}; quadratic fits over increasing neighborhoods give
\SIrange{0.068}{0.073}{\micrometre}. The two solvers differ in the absolute common emission-axis
coordinate by approximately \SI{1.3}{\micrometre}, but both move the wavelengths together and
agree on their relative registration and spot shape. The absolute offset is a mask-to-ion placement
coordinate that is calibrated after fabrication; the fixed-layout prediction is the relative
two-color focus.

\begin{figure}[htbp]
\centering
\includegraphics[width=0.96\linewidth]{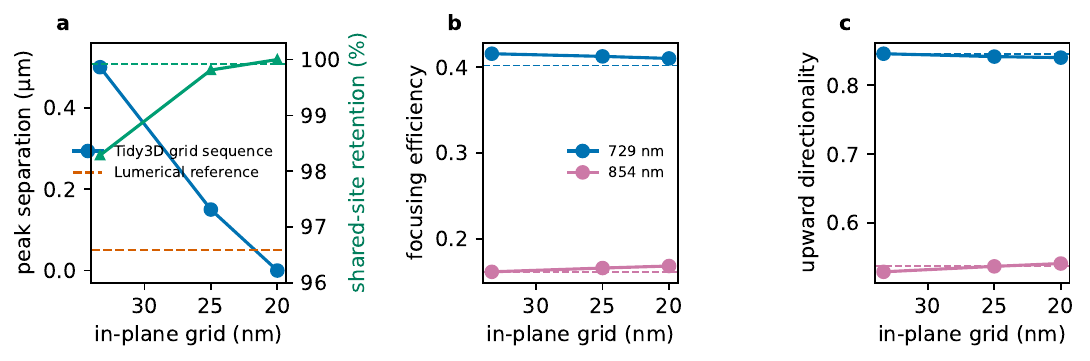}
\caption{\textbf{Numerical refinement and independent validation of the registered design.}
(a) Tidy3D peak separation and shared-site retention versus lateral mesh. (b) Focusing efficiency
and (c) directionality for both channels. Circles and solid curves show the controlled Tidy3D
lateral-grid sequence; horizontal dashed lines show the independently built Lumerical calculation,
whose lateral mesh is automatic rather than assigned a nominal spacing.}
\label{fig:validation}
\end{figure}

The finite flux monitors above, below, and at the lateral boundaries overlap and do not form a
single nonoverlapping closed surface. Their sums are therefore coverage checks, not strict energy
closures. For the final Tidy3D calculation the all-side coverage sums are 0.978 and 0.995 of input
power; the corresponding Lumerical up+down+forward sums are 0.984 and 0.988. Guided-mode forward
fractions at the input exceed 0.987 in Tidy3D, the computed TE fraction is unity to numerical
precision, and the reflected-to-forward guided power summed over the four recorded modes is
$2.4\times10^{-3}$ at \SI{729}{\nano\meter} and $1.3\times10^{-3}$ at
\SI{854}{\nano\meter}.

The radiated fraction $\eta_{\mathrm{rad}}=\eta_{\uparrow}/D$ is 0.846 at 729 nm and 0.704 at
\SI{854}{\nano\meter}. At 854 nm, the forward output cap records 0.209 of the input power and the
two lateral caps together record 0.082; the corresponding 729-nm fractions are 0.075 and 0.056.
Because these finite monitors overlap in coverage, the cap values describe escape channels but are
not summed as an exact power partition. Together with $D=0.541$ at 854 nm, they show that incomplete
aperture extraction and downward radiation both reduce $\eta_{\uparrow}$.

\section{Context against trap-integrated emitters}

Table~\ref{tab:prior-art} places the reported numbers beside representative trap-integrated and
simulation studies. Each entry preserves the metric defined by its source because spot size,
crosstalk, efficiency, and experimental control error are distinct observables. The comparison
isolates the contribution here---independent
narrowband wavefronts stacked in one aperture---from records obtained for single-color or shared
broadband optics.

\begin{table}[htbp]
\centering
\caption{Representative integrated trapped-ion light-delivery results. ``Sim.'' denotes a numerical
study and ``exp.'' an experimental device. Values retain the metric used by each source.}
\label{tab:prior-art}
\small
\resizebox{\linewidth}{!}{%
\begin{tabular}{lllll}
\toprule
Work & architecture/status & height & reported spatial metric & reported efficiency metric \\
\midrule
Mehta et al.~\cite{Mehta2016} & single-color grating, exp. & 50 $\mu$m & 2 $\mu$m $1/e^2$ radius; error $10^{-2}$--$4\times10^{-4}$ at 7.5--15 $\mu$m & not reported in common FE form \\
Niffenegger et al.~\cite{Niffenegger2020} & multiple couplers, violet--infrared, exp. & --- & multicolor ion control & wavelength-dependent insertion loss \\
Shirao et al.~\cite{Shirao2022} & double-etched SiN grating, sim. & --- & 2.5 $\mu$m diameter; $-36$ dB at 7.07 $\mu$m & 98\% directionality \\
Momenzadeh et al.~\cite{Momenzadeh2026} & multimode grating circuit, sim. & 80 $\mu$m & point sample: $-21.1$ dB at 5 $\mu$m & FE $-3.8$ dB \\
Klawson et al.~\cite{Klawson2026} & broadband lens/micromirror, exp. & 75 $\mu$m & average $-27$ dB at 5 $\mu$m & 405--880 nm operation \\
This work & two stacked narrowband gratings, sim. & 70 $\mu$m & point/disk XT $-24.8/-18.0$ and $-15.7/-13.3$ dB at 5 $\mu$m & FE 0.410/0.168 \\
\bottomrule
\end{tabular}}
\end{table}

\section{Near-field propagation and focusing efficiency}

The complex vector field is recorded on a plane \SI{0.5}{\micrometre} above the top cladding and
propagated by the band-limited angular-spectrum method~\cite{Matsushima2009}. For each transverse
spatial frequency $(k_x,k_y)$,
\begin{equation}
\widetilde{\mathbf E}(k_x,k_y;z)=\widetilde{\mathbf E}(k_x,k_y;0)
\exp\!\left[i z\sqrt{k_0^2-k_x^2-k_y^2}\right],
\end{equation}
with evanescent components excluded and the distance-dependent Matsushima--Shimobaba band limit
applied. Before propagation, the same super-Gaussian window with field exponent 16---equivalently,
order 8 in intensity---and
\SI{11}{\micrometre} half-width is applied to both solvers to remove fields from the mode-launch
junction outside the patterned aperture.
The propagated analysis plane is $[-44,44]\times[-44,44]~\si{\micrometre^2}$ on a common
\SI{50}{\nano\meter} reporting grid. The production conditioner retains 82.7\% and 70.3\% of the
sampled near-field $|E|^2$ proxy at 729 and \SI{854}{\nano\meter}, respectively.

The upward coupling is $\eta_\uparrow=P_\uparrow/P_\mathrm{in}$ and directionality is
$D=P_\uparrow/(P_\uparrow+P_\downarrow)$. The clean focal fraction is
\begin{equation}
f_{2\mu\mathrm{m}}=
\frac{\int_{|\mathbf r-\mathbf r_0|\leq\SI{2}{\micrometre}}
  (|E_x|^2+|E_y|^2+|E_z|^2)\,dA}
{\int_{[-44,44]^2\,\si{\micrometre^2}}(|E_x|^2+|E_y|^2+|E_z|^2)\,dA},
\end{equation}
and $\mathrm{FE}=\eta_\uparrow f_{2\mu\mathrm{m}}$. This flux-times-focal-fraction metric uses the
same guided-input normalization for every quoted value and both solvers.

Table~\ref{tab:conditioning} reports final-mask sensitivity to field conditioning. The unwindowed and alternate
windows change the focusing efficiencies by at most 0.017 absolute, keep shared-site retention above
99\%, and preserve the crosstalk ordering. Replacing the $|E|^2$ focal fraction by the forward
Poynting-flux fraction changes FE by +0.21\% at 729 nm and $-0.74$\% at 854 nm.

\begin{table}[htbp]
\centering
\caption{Final-mask sensitivity to near-field conditioning. Paired entries are 729/854 nm. The
super-Gaussian exponent is the field exponent.}
\label{tab:conditioning}
\small
\resizebox{\linewidth}{!}{%
\begin{tabular}{lcccccc}
\toprule
Conditioner & near-field $|E|^2$ retained & FE & shared site (\%) & peak sep. (\si{\micrometre}) & point XT 729/854 (dB) & disk XT 729/854 (dB) \\
\midrule
super-Gaussian, 11 $\mu$m, $p=16$ & 0.827/0.703 & 0.410/0.168 & 100.00 & 0.00 & $-24.84/-15.73$ & $-18.03/-13.31$ \\
super-Gaussian, 10 $\mu$m, $p=16$ & 0.703/0.584 & 0.413/0.169 & 99.75 & 0.15 & $-19.90/-14.05$ & $-15.36/-12.99$ \\
super-Gaussian, 12 $\mu$m, $p=16$ & 0.903/0.809 & 0.404/0.162 & 99.58 & 0.20 & $-23.71/-16.89$ & $-18.81/-13.71$ \\
super-Gaussian, 13.5 $\mu$m, $p=16$ & 0.961/0.922 & 0.395/0.156 & 99.29 & 0.30 & $-22.11/-16.99$ & $-17.91/-13.76$ \\
hard square, 12 $\mu$m & 0.956/0.899 & 0.397/0.155 & 99.38 & 0.30 & $-22.11/-17.48$ & $-17.89/-13.93$ \\
no conditioner & 1.000/1.000 & 0.393/0.154 & 99.14 & 0.35 & $-21.70/-16.84$ & $-17.72/-13.69$ \\
\bottomrule
\end{tabular}}
\end{table}

The propagation implementation was independently tested with a scalar Gaussian field of
$1/e^2$ intensity radius \SI{6}{\micrometre}. After \SI{67.5}{\micrometre}, the numerical and
analytic paraxial radii are 6.570/\SI{6.544}{\micrometre} at 729 nm and
6.712/\SI{6.735}{\micrometre} at 854 nm. The relative radius errors are +0.41\% and $-0.34$\%; the
on-axis intensity errors are $-0.05$\% and $-0.07$\%. This analytic check is independent of the
Tidy3D--Lumerical near-field comparison.

\begin{figure}[htbp]
\centering
\includegraphics[width=0.92\linewidth]{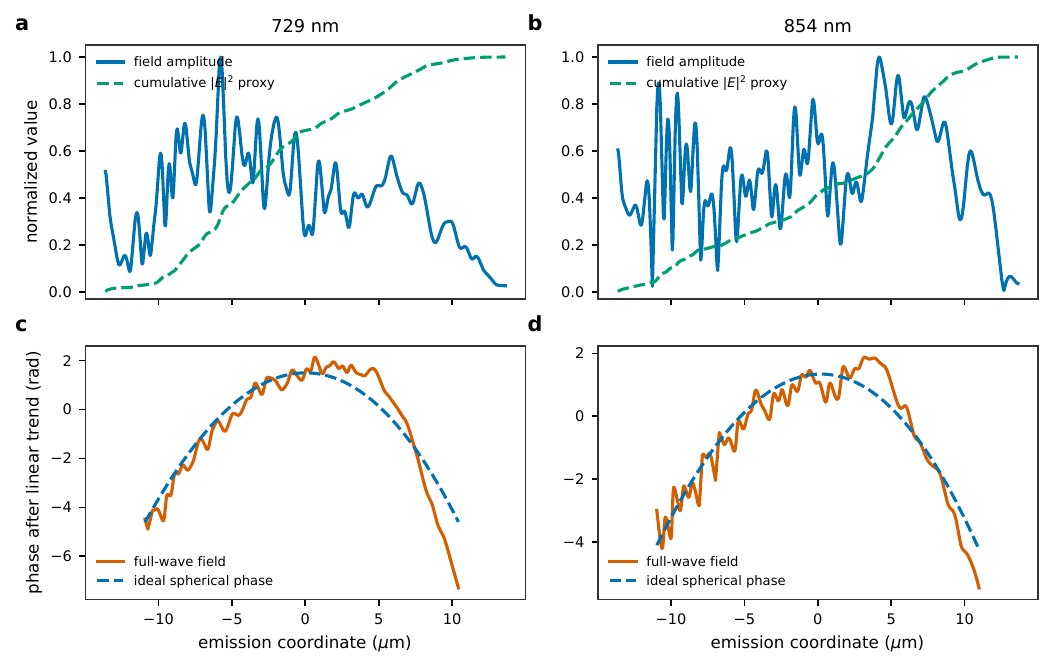}
\caption{\textbf{Emitted aperture field.} (a,b) Centerline field amplitude and cumulative
$|E|^2$ proxy on the final near-field plane. (c,d) Unwrapped phase of the dominant transverse
component after a \SI{0.2}{\micrometre} complex-envelope smoothing and removal of the fitted carrier
slope. Dashed curves are the ideal spherical phase converging on the shared site; phase is shown
where the smoothed amplitude exceeds 20\% of its maximum.}
\label{fig:aperture-field}
\end{figure}

Figure~\ref{fig:aperture-field} exposes the emitted-field mechanism directly. The centerline
amplitude is progressively accumulated across each aperture, while the unwrapped phase of the
dominant transverse component follows the converging spherical-wave target after removal of the
carrier slope. The cumulative $|E|^2$ trace is a near-field radiation proxy; longitudinal guided
power is quantified separately by the flux monitors.

The shared site maximizes the smaller of the two individually normalized intensities,
\begin{equation}
\mathbf r_s=\underset{\mathbf r}{\arg\max}\;
\min\!\left[\frac{I_{729}(\mathbf r)}{I_{729}^{\max}},
            \frac{I_{854}(\mathbf r)}{I_{854}^{\max}}\right].
\end{equation}
Shared-site retention is the value of this maximum. Unlike an argmax separation, it directly states
how much of each color is available at one physical ion coordinate.

\section{Ion-height operating range}

The final Tidy3D near fields were repropagated over a range of ion heights without another FDTD
solve. Height is measured from the top surface of the upper SiN film; because the nominal optical
overclad is \SI{2}{\micrometre} thick, clearance from its outer surface is \SI{2}{\micrometre}
smaller. At each height the shared coordinate was reselected using the definition above, and
crosstalk was sampled along the ion-chain axis. Table~\ref{tab:height} shows representative planes.

\begin{table}[htbp]
\centering
\caption{Registered-device height scan. Crosstalk is evaluated at the common site and its
$\pm\SI{5}{\micrometre}$ chain neighbors.}
\label{tab:height}
\small
\resizebox{\linewidth}{!}{%
\begin{tabular}{cccccc}
\toprule
Height (\si{\micrometre}) & Shared site (\%) & Peak sep. (\si{\micrometre}) &
729 point/disk (dB) & 854 point/disk (dB) & Chain FWHM 729/854 (\si{\micrometre}) \\
\midrule
60.0 & 99.00 & 0.25 & $-14.9/-14.8$ & $-16.4/-14.0$ & 2.05/2.55 \\
65.0 & 99.84 & 0.10 & $-19.5/-16.9$ & $-17.3/-13.5$ & 2.15/2.55 \\
67.5 & 99.98 & 0.05 & $-22.8/-18.6$ & $-16.5/-13.3$ & 2.15/2.60 \\
70.0 & 100.00 & 0.00 & $-24.8/-18.0$ & $-15.7/-13.3$ & 2.25/2.65 \\
72.13 & 99.93 & 0.05 & $-23.6/-16.8$ & $-15.3/-13.4$ & 2.25/2.75 \\
72.5 & 99.85 & 0.15 & $-23.1/-16.6$ & $-15.1/-13.4$ & 2.25/2.75 \\
75.0 & 99.79 & 0.20 & $-20.5/-15.4$ & $-14.7/-13.7$ & 2.35/2.80 \\
85.0 & 98.73 & 0.50 & $-14.4/-12.8$ & $-14.5/-14.4$ & 2.70/3.15 \\
\bottomrule
\end{tabular}
}
\end{table}

\begin{figure}[htbp]
\centering
\includegraphics[width=0.94\linewidth]{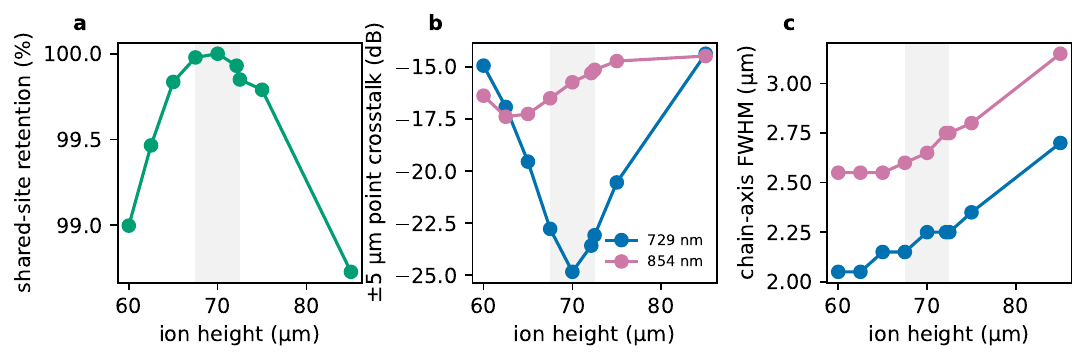}
\caption{\textbf{Ion-height operating range.} Shared-site retention, chain-axis FWHM, and
nearest-neighbor point crosstalk obtained by propagating the same finest-mesh near fields. The
selected \SI{70}{\micrometre} plane maximizes registration and minimizes 729-nm point crosstalk.}
\label{fig:height}
\end{figure}

\section{Crosstalk and quadrupole coupling}

Point crosstalk is
$10\log_{10}[I(\mathbf r_s\pm\SI{5}{\micrometre}\,\hat{y})/I(\mathbf r_s)]$.
The single value reported for each wavelength is the larger (less negative) of the two
$+\SI{5}{\micrometre}$ and $-\SI{5}{\micrometre}$ neighbor ratios.
The disk metric is the maximum over a \SI{0.5}{\micrometre}-radius neighborhood of the nominal
neighbor. It represents a conservative combination of ion-position and beam-pointing uncertainty,
rather than an ion-size model. Values at \SI{10}{\micrometre} are visible in the main-text
profiles but are not used as the headline metric.

For orientation, $-24.84$~dB intensity crosstalk corresponds to a field-amplitude ratio of 0.057.
If both ions were exactly resonant and a rectangular target $\pi$ pulse acted on the neighbor with
that reduced Rabi frequency, the spectator excitation would be
$\sin^2[(\pi/2)\sqrt{10^{-24.84/10}}]\approx8.1\times10^{-3}$. Applying the same idealized
calculation to the $-18.03$~dB disk value gives approximately $3.8\times10^{-2}$. These are
reference values for a bare resonant pulse. Detuning, Zeeman selection, pulse shaping, echo or
composite sequences, and the actual quadrupole tensor determine the experimental gate error.

For the \SI{729}{\nano\meter} electric-quadrupole transition, the optical field gradient rather
than intensity alone sets the Rabi frequency,
\begin{equation}
\Omega\propto\sum_{ij}\langle D,m'|Q_{ij}|S,m\rangle\,\partial_iE_j.
\end{equation}
The exported field was additionally reprocessed in free space to obtain
$G=\|\operatorname{sym}_0(\nabla\mathbf E)\|_F^2$, the squared Frobenius norm of the symmetric,
traceless field-gradient tensor. Spatial derivatives in the ion plane were evaluated on the common
\SI{50}{\nano\meter} reporting grid; the normal derivative was obtained spectrally from the same
angular-spectrum representation. The field at the shared site is more than 99.99\% in the dominant
transverse component in both channels. At the two \SI{5}{\micrometre} chain neighbors, the worst
ratios $G/G_0$ are $-23.0$~dB at \SI{729}{\nano\meter} and $-15.6$~dB at
\SI{854}{\nano\meter}. This rotationally invariant norm tests whether spatial discrimination
survives a vector-gradient description; contraction with $Q_{ij}$ gives the coupling for a selected
Zeeman line.

Consequently, FE, scalar crosstalk, and $G/G_0$ remain electromagnetic delivery metrics. A
transition-specific gate-error calculation requires a Zeeman line, magnetic-field axis, polarization
at the ion, pulse envelope, and spectator detuning. For the allowed \SI{854}{\nano\meter}
transition, the point and disk intensities at a \SI{5}{\micrometre} neighbor are 2.7\% and 4.7\% of
the target value. The upper channel is consequently designed for shared or global repumping; a
site-selective reset variant would add neighbor suppression to the optimization objective.

\section{Minimum-feature and comparator definitions}

The \SI{125}{\nano\meter} propagation-axis feature rule is evaluated by scanning each 2D height
map along the local propagation direction. Contiguous line and space runs are measured on the \SI{62.5}{\nano\meter}
design raster. A two-cell threshold therefore corresponds to \SI{125}{\nano\meter}. The local
grating period is used when short runs are repaired, preventing a period chirp from reintroducing an
undersized feature near the aperture edge. The final depth-allocated design has no sub-threshold
propagation-axis run under this procedure.

The vector GDS was also decomposed into the four physical sequential etch masks and screened by
erosion/opening operations. The large line/space-loss onset occurs at the encoded
\SI{125}{\nano\meter} threshold. Curved staircase boundaries retain
\SI{62.5}{\nano\meter} jogs; the stated feature floor applies to propagation-axis line and space.
Foundry submission then adds the process-design-kit rule deck, bias and corner rounding, overlay
tolerances, and lithographic process model.

The \SI{729}{\nano\meter} width-taper reference uses a related aperture but permits
\SI{62.5}{\nano\meter} runs and was optimized for a different lateral phase target before the final
registration step. At \SI{20}{\nano\meter} lateral resolution it gives FE 0.387, chain FWHM
\SI{2.25}{\micrometre}, point crosstalk $-22.9$~dB, and disk crosstalk $-19.1$~dB. The associated
depth-allocated precursor gives 0.398, \SI{2.25}{\micrometre}, $-24.7$~dB, and $-18.0$~dB under
the same analysis. Because the targets and allowed feature sets differ, this is a mechanism and
fabricability diagnostic rather than a constraint-matched performance ranking. The result used in
the main text is solely geometric: the depth-allocated final mask removes the sub-threshold
propagation-axis runs present in this width-taper construction.

\section{Model parameters and fabrication transfer}

Table~\ref{tab:scope} summarizes how the optical model connects to a fabrication process. The final
two-solver calculations use vertical sidewalls and the same real,
wavelength-independent indices, $n_{\mathrm{SiN}}=2.018$ and $n_{\mathrm{SiO_2}}=1.453$.
Agreement between those calculations establishes the nominal mask response. Process transfer adds
dispersion, absorption, sidewall taper, and roughness from ellipsometry and cross-sectional
metrology of the selected films and etch process.

\begin{table}[htbp]
\centering
\caption{Numerical treatment and process-transfer inputs for the final registered mask.}
\label{tab:scope}
\footnotesize
\resizebox{\linewidth}{!}{%
\begin{tabular}{lll}
\toprule
Quantity & Treatment in this work & Role in process transfer \\
\midrule
Lateral and vertical mesh & Tidy3D lateral sequence; independent Lumerical mesh & Fixed-mask optical response \\
Upper phase target & Refined through full-stack solves & Two-color focal registration \\
Relative focal displacement & Translated-field response & 98\% shared-site margin \\
Material indices & Real values at two design lines & Replace with film ellipsometry \\
Etch depth and film thickness & Nominal mask stack & Map measured process corners \\
Sidewall geometry & Vertical interfaces & Insert measured etch profile \\
Interlayer overlay & Mask-registered nominal stack & Apply lithographic overlay statistics \\
External feed network & Single active input per solve & Add routing-specific ports and bends \\
\bottomrule
\end{tabular}
}
\end{table}

Four sequential etches are implied by the two depth levels in each of the two films. The
\SI{125}{\nano\meter} result is the deterministic propagation-axis line-and-space constraint.
Interlayer overlay, thickness, and sidewall perturbations alter each color's wavefront, while the two
colors remain mutually incoherent. Statistical yield analysis follows by sampling the correlated
distributions, etch biases, dispersive indices, and sidewall statistics of the selected process from
the released GDS and solver inputs.

\section{Design-stage FDTDX model and resolution limit}

FDTDX provided a differentiable single-film, mirror-free unit-cell model and the initial
full-aperture design environment. Its completed
final-mask calculation used a \SI{62.5}{\nano\meter} isotropic grid. The physical DBR layers, with thicknesses of \SI{86.72}{\nano\meter} and
\SI{120.44}{\nano\meter}, span only 1.39 and 1.93 cells and are rounded by the Yee
representation. The differentiable cell study therefore supplied the starting local scatterer and
initial layout, while the final numerical values come from the laterally refined Tidy3D model and
the independently built Lumerical model.

Reflector spacing is a primary efficiency-determining parameter because it sets the round-trip
phase between the aperture and the buried reflector. It must therefore be co-targeted with the
calibrated oxide thickness of the selected fabrication process. The translated-field study in
Figure~\ref{fig:registration-margin} independently gives the focal-overlap margin of the final mask.

\section{Electrode-footprint placement calculation}

The placement calculation uses the gapless surface-electrode boundary solution of
House~\cite{House2008}. The \SI{24}{\micrometre} optical aperture is translated across the grounded
central region toward an RF rail, and the radial RF-null displacement is solved at each position.
The unperturbed House null lies \SI{70.13}{\micrometre} above its conductor plane. If that plane is
placed at the outer surface of the nominal \SI{2}{\micrometre} overclad, the House height corresponds
to an optical height of \SI{72.13}{\micrometre} above the upper film. Repropagation of the stored
finest-mesh near fields to that exact plane gives the common coordinate
$(24.225,-0.025)~\si{\micrometre}$, 99.93\% shared-site retention, and 729-nm point and disk
crosstalk of $-23.58$ and $-16.80$~dB. Placing the aperture center at
$-\SI{24.225}{\micrometre}$ therefore aligns the common optical coordinate laterally with the
symmetry line.
The inner RF-rail boundary is at $-\SI{51.5}{\micrometre}$ and the nearest aperture edge is at
$-\SI{36.225}{\micrometre}$, giving \SI{15.275}{\micrometre} clearance. An aperture center at
$-\SI{39.5}{\micrometre}$ is the first-contact threshold, not the rail coordinate.

Within the House model, an aperture wholly inside the central grounded region is assigned exactly
the same Dirichlet boundary condition as the ground plane. The zero displacement therefore follows
by construction and establishes that the scaled footprint can be positioned without cutting the
idealized RF rail. Device-level trap analysis then introduces finite electrode gaps and thickness,
dielectric permittivity and charge, secular confinement, and transparent conductive shielding in an
implementation-specific electrostatic/RF model.

\bibliographystyle{achemso}
\bibliography{references}